\documentclass{ajour}
\usepackage{epsfig}

\begin{document}

\authorrunninghead{Pradosh Mohapatra}
\titlerunninghead{Fully Dynamic Minimum Spanning Tree Algorithms}

\title{Fully Sequential and Distributed Dynamic Algorithms for Minimum Spanning Trees}


\author{Pradosh Kumar Mohapatra}

\affil{Department of Electrical Engineering and Computer Science, \\University of Illinois at Chicago, \\Chicago, Illinois - 60607, USA}

\email{pmohapat@eecs.uic.edu}

\abstract{In this paper, we present a fully-dynamic distributed algorithm for maintaining a minimum spanning tree on general graphs with positive real edge weights. The goal of a dynamic MST algorithm is to update efficiently the minimum spanning tree after dynamic changes like edge weight changes, rather than having to recompute it from scatch each time. The first part of the paper surveys various algorithms available today both in sequential and distributed environments to solve static MST problem. We also present some of the efficient sequential algorithms for computing dynamic MST like the Frederickson's algorithm and Eppstein's sparsification technique. Lastly we present our new sequential and distributed algorithms for dynamic MST problem. To our knowledge, this is the first of the distributed algorithms for computing dynamic MSTs.
}
\baselineskip=1.5em
\begin{article}
\section{Introduction} \label{introduction}
The {\it minimum spanning tree} or $MST$ problem is one of the simplest and best-studied optimization problems in computer science. Given an undirected, connected graph $G$ with $n$ vertices and $m$ weighted edges, the $MST$ problem is to find a spanning tree, i.e., a tree that connects all the vertices of $G$ using only edges of $G$ of minimum total weight. The history of $MST$-construction algorithms goes way back to early nineteenth century (Bor\.{u}vka in 1926 and Varn\'{i}k in 1930). But the most famous and classical algorithms studied today are the 1950's $MST$ algorithms by Kruskal~\cite{CLR} and Prim~\cite{CLR}.
\par \indent Spanning trees are also essential in distributed computation. Processes are connected through an arbitrary communication network, which is essentially a graph. Spanning trees can enforce synchronization on this network and thus enable centralized applications to run on asynchronous networks. They are needed in synchronization protocols and distributed algorithms, such as breadth-first search. We can also think of problems in distributed databases as deadlock resolution, or the replacement of a malfunctioning central lock-coordinator. In all these cases, instead of using an arbitrary spanning tree for the broadcast of messages, we would prefer an $MST$ that minimizes some cost function. Several distributed algorithms have been presented, the pioneering one being ~\cite{Gallager}.
\par \indent While this "easy" optimization problem is an interesting area to look into, there is another more interesting field associated with it that has gained wide attention and interest in the last decade, namely dynamic $MST$ problem.In many applications of the $MST$, including communication networks, assembly planning, and VLSI design, the underlying graphs are subject to discrete changes, such as additions or deletions of edges and/or vertices. The goal of the dynamic $MST$ algorithm is to update efficiently the minimum spanning tree after dynamic changes, rather than having to recompute it from scratch each time. Given its powerful versatility, it is not surprising that dynamic $MST$ algorithms and the associated data structures are more difficult to design and analyze than the static counterpart.
\par \indent For the sequential dynamic $MST$ problem, there have been considerable progress over the last decade. The most prominent among them are (1) dynamic trees data structure of Sleator and Tarjan~\cite{Sleator}, (2) topology trees data structure of Frederickson~\cite{Frederickson}, and (3) Sparsification technique suggested by Eppstein~\cite{Eppstein}. While the dynamic tree solution is very efficient in maintaining the $MST$ in case of edge deletions(a non-tree edge decreasing weight as will be seen later), topology trees are more general and solve both edge insertions and deletions. The sparsification technique is a {\it black box} approach to reduce the time complexity of any such algorithm.
\par \indent While the sequential dynamic $MST$ problem is a rich subject, to our knowledge, there have not been many distributed algorithms for the dynamic $MST$ problem till date. Pawagi and Ramakrishnan ~\cite{Pawagi} were the first ones to give a parallel algorithm to the problem which runs in $O(\log n)$ time using $n^2$ $CREW$ $PRAM$s. The main aim of this term paper is to shed some light towards the dynamic computation of $MST$s in distributed networks. Initially, we give an efficient algorithm for the sequential dynamic MST problem based on some important properties of graphs and MSTs that we have found. Then, starting with an attempt to convert the sequential dynamic algorithms to distributed algorithms, we present new techniques for the problem.
\par \indent The rest of the document is organized as follows. In chapter 2, we provide the background and a set of preliminary definitions for the general $MST$. Chapter 3 discusses the early sequential algorithms to computer $MST$s, namely the well-known Prim's and Kruskal's algorithms. In chapter 4, we study the distributed algorithms for computing $MST$s: the first efficient algorithm by Gallager, Humblet, and Spira ~\cite{Gallager}, node counting improvement by Chin and Ting ~\cite{Chin} and the optimal algorithm by Awerbuch ~\cite{Awerbuch}. In chapter 5, we explain sequential dynamic MST algorithms. Starting with Frederickson's topology tree ~\cite{Frederickson} description and Eppstein's sparsification technique ~\cite{Eppstein}, we give our version of the sequential MST algorithm. Chapter 6 explores the distributed dynamic MST problem. We first give a brief idea of the parallel algorithm developed by Pawagi and Ramakrishnan ~\cite{Pawagi} for solving the problem, and then outline the algorithm steps for efficiently computing  $MST$ dynamically in a distributed system. We conclude the paper with chapter 7. 
\section{Background and Preliminary Definitions}
\subsection{Notations and Conventions}
$G\ =\ (V,E)$ denotes the undirected input graph, and $M\ =\ (V,E')$ denotes the correct minimum spanning tree. $G$ has $n$ vertices, and $m$ edges. We standardly assume that $G$ is connected, so $m\ \geq\ n-1$. An edge $e\ \in\ E$ has weight $w(e)$. If $e$'s endpoints in the graph are $u$ and $v$, we may sometimes denote it by $(u,v)$.
For simplification, we assume the edge weights are distinct real numbers. 
\subsection{Graph Theory Definitions}
A {\bf {\it cut}}$(S,V-S)$ of an undirected graph $G(V,E)$ is a partition of $V$ into two sets $S$ and $V-S$. We say that an edge $(u,v)\ \in\ E$ {\bf {\it crosses}} the cut $(S,V-S)$ if one of its endpoints is in $S$ and the other is in $V-S$. An edge $e$ is a {\bf {\it light edge}} across a cut if its weight is the minimum of any edge crossing the cut. 
\subsection{Properties of Spanning Trees}
A spanning tree, $T$, is defined as a connected acyclic spanning subgraph of $G$. {\it Connected} means that $T$ includes at least on edge crossing each cut of $G$. {\it Acyclic} means that $T$ excludes at least one edge from each cycle of $G$. A minimum spanning tree, $M$, is a spanning tree of $G$ whose edges have minimal total weight. We will use the notation $w(M)$ to denote this total weight.
\subsection{Properties of the MST}
Under our assumption that all edges have distinct weights, the minimum spanning tree $M$ has the following well-known complementary properties:
\begin{itemize}
\item {\bf Strong Cut Property}: $e$ $\in\ M\ \Leftrightarrow$ $e$ is the lightest edge across some cut of $G$.
\item {\bf Strong Cycle Property}: $e$ $\notin\ M\ \Leftrightarrow$ $e$ is the heaviest edge on some cycle of $G$.
\end{itemize}
Either property implies at once that $M$ is unique.
\section{Sequential MST Algorithms}
All sequential MST algorithms are rapid methods for ruling edges in or out of $M$. We will consider two such classical algorithms, Kruskal's algorithm and Prim's algorithm. As we will see, these two algorithms are instances of a "generalized greedy algorithm" for constructing $MST$s. It initializes a forest $F$ to $(V, \emptyset)$, and adds edges one at a time till $F$ is connected. Every edge that it adds is the lightest edge leaving some component $T$ of $F$. 
\subsection{Kruskal's Algorithm}
This classical algorithm derives directly from the properties of the $MST$ discussed above. We consider each edge $e$, and use the cut property and cycle property to decide correctly whether $e\ \in\ M$. If so, we add it to a growing forest $F\ \subseteq\ M$; if not, we discard it. More formally, the following is the algorithm:
\begin{algorithm}[MST-Kruskal$(G)$]
\   begin
\      $F\ \leftarrow\ \emptyset$  /* the growing forest which will eventually be M */
\      $A\ \leftarrow\ \emptyset$  /* for cycle detection, usually implemented by the union-find algorithm */
\      sort the edges of $E$ by nondecreasing weight $w$.
\      for each edge $(u,v)\ \in\ E$, in order of nondecreasing weight.
\         if $u$ and $v$ both do not belong to $A$
\            $F\ \leftarrow\ F\ \cup\ \{(u,v)\}$
\            $A\ \leftarrow\ A\ \cup\ \{u\}\ \cup\ \{v\}$
\         end if.
\      end for.
\   end.
\end{algorithm}
\subsubsection{Analysis}
The sorting procedure in line $4$ takes $O(m \log m)$. The other main part is the loop at lines $5-9$. There are $m$ iterations of the loop, and at each iteration, the main procedure is to determine for arbitrary edge $(u,v)$ whether there is already a $u\ \ldots\ v$ path in $F$. Using the union-find algorithm, this takes $\alpha(m,n)$ time where $\alpha$ is the functional inverse of Ackermann's function. Total time in the loop at lines $5-9$ is $O(m \alpha(m,n)$. 
\par \indent Since $\alpha(m,n)\ =\ O(\log m)$, the total running time of Kruskal's algorithm is $O(m \log m)$.
\subsection{Prim's Algorithm}
Like Kruskal's, Prim's algorithm is also a special case of the generic greedy algorithm. However, rather than growing many trees simultaneously, it devotes all its energy to growing a single tree $T$ of the forest $F$.
\begin{algorithm}[MST-Prim$(G)$]
\   begin
\      $F\ \leftarrow\ \emptyset$  /* the growing forest which will eventually be M */
\      $u\ \leftarrow\ $an arbitrary start vertex in $V$.
\      repeat $(n-1)$ times
\         $e\ \leftarrow\ $ the lightest edge of $G$ leaving $F$. (i.e. having just one endpoint in $F$).
\         $F\ =\ F\ \cup\ \{e\}$
\      end repeat
\   end
\end{algorithm}
\subsubsection{Implementation and Analysis}
The key to implementing Prim's algorithm efficiently is to make it easy to find the lightest edge leaving $F$. In a typical implementation, all vertices {\it not} in the tree reside in a priority queue $Q$ based on a {\it key} field. On every iteration, we $EXTRACT-MIN$ the vertex $u$ that is closest to $F$ and add the lightest edge from $u$ to $F$. When the algorithm terminates, the priority queue $Q$ is empty; and $F$ is the $MST$ for the graph.
\par \indent So the performance of Prim's algorithm depends on how we implement the priority queue $Q$. The asymptotic running time comes to be the best when it is implemented by Fibonacci heaps. In a Fibonacci heap, we can perform the $EXTRACT-MIN$ operation in $O(\log n)$ amortized time. It can be verfied easily that using Fibonacci heaps, the total running time of Prim's algorithm comes to $O(m + n \log n)$.
\section{Distributed MST Algorithms}
The study of algorithms for a sequential computer has been a highly successful endeavor, providing a common framework for devising algorithms and comparing their performance. The goal of distributed algorithms, or distributed computing in a general sense, is to accomplish the same for distributed systems. Unfortunately, because of the wide differences between systems, there is not a universally accepted model of computation. Nonetheless, since the late 70s, there has been intensive research in applying various theoretical paradigms to distributed systems. As our main aim is to devise and analyze distributed dynamic algorithms, we will discuss more on the distributed static $MST$ algorithms than we did for their sequential counterparts. 
\subsection{Model of Computation}
We consider message passing systems with no failures. In a message passing system, processors communicate by sending messages over communication channels, where each channel provides a bidirectional connection between two specific processors. The pattern of connections provided by the channels describes the {\it topology} of the system. The topology is represented by an undirected graph in which each node represents a processor, and an edge is present between two nodes if and only if there is a channel between the corresponding processors. Furthermore, we assume our timing model to be asynchronous, i.e. there is no fixed bound on how long it takes for a message to be delivered or how much time elapses between consecutive steps of a processor. An algorithm for such a message passing system consists of a local program for each processor in the system. It provides the ability for the processor to perform local computation and to send messages to and receive messages from each of its neighbors in the given topology.
\subsection{Complexity Measures}
We will consider two complexity measures: {\it message complexity}, and {\it time complexity}. The message complexity of an algorithm is the meximum, over all {admissible executions}~\cite{Attiya} of the algorithm, of the total number of messages sent. As there is no bound on the message delay in an asynchronous model, we will assume that the maximum message delay in any execution is one unit of time and define the time complexity as the maximum time until termination among all {\it timed admissible executions}~\cite{Attiya} assuming the above.
\subsection{Distributed MST Problem Model}
The distributed system, as described above, is modeled by an arbitrary undirected, connected topology graph $G(V,E)$ with $n$ nodes and $m$ edges. With each edge $e\ \in\ E$, we associate a weight $w(e)$, a unique real number. The system is asynchronous and we further assume that the edges follow a FIFO policy, i.e. messages arrive in the order they were sent. At the beginning of the algorithm, a processor knows only the weights of the edges adjacent to it. Processors start the algorithm either spontaneously or upon receiving a message from a neighbor. It is necessary to assume that either edges have distinct weights or nodes have unique id's. Otherwise, there is no distributed algorithm for finding an $MST$, because in that case, it is similar to a non-uniform anonymous algorithm for leader election in an asynchronous rings; which has an impossiblity result~\cite{Attiya}.
\subsection{Preliminaries}
Let a {\it fragment} of an $MST$ be a connected subgraph of it. An {\it outgoing edge} of a fragment is an edge with one adjacent node in the fragment and the other adjacent node not in the fragment. Define the {\it minimum outgoing edge(MOE)} of a fragment to be the outgoing edge of the fragment with minimum weight.
\begin{lemma} \label{fragmentlemma}
Let $G(V,E)$ be a connected graph with distinct weights. Let $M$ be its unique $MST$. For any fragment $F$ of $M$, the $MOE$ of $F$ is in $M$.
\end{lemma}
{\bf {\it Proof:}} By contradiction. Let $e$ be the $MOE$ of $F$. Then $w(e)\ <\ w(e')$ for any other $e'\ \in\ F$. Assume that $e\ \notin\ M$. Then $M\ \cup\ \{e\}$ contains a cycle (addition of an additional edge to a tree). This cycle contains $e$ and at least one additional edge of the fragment $F$, say $e'$. Then $M\ \cup\ \{e\}\ \setminus\ \{e'\}$ forms a spanning tree and $w(M\ \cup\ \{e\}\ \setminus\ \{e'\})\  <\  w(M)$. That defies the fact that $M$ is the unique $MST$. A contradiction.
\par \indent All the distributed algorithms proposed so far for the $MST$ problem have the same general structure as the sequential algorithms. Not surprising, we can, in the same way, define a "generic distributed MST algorithm":At the beginning, each node is a separate fragment. In each stage of the algorithm, each fragment finds its $MOE$, and attempts to combine with the fragment at the other end of the edge. By ~\ref{fragmentlemma}, such a combination yields a new bigger fragment of the $MST$. The algorithm ends when there is only one fragment, which is the $MST$. It differs from its sequential counterpart in the parallelism of the fragments' combinations. 
\subsection{The Pioneering Work of Gallager, Humblet, and Spira ~\cite{Gallager}}
The basic steps of the algorithm are as follows:
\begin{enumerate}
\item Each process starts the algorithm as an individual fragment.
\item After a new fragment is created, it chooses its $MOE$.
\item The fragment tries to join with the fragment at the other end of the chosen $MOE$ to form a bigger fragment.
\item This process continues till there can no more be $MOE$s chosen, which means that there is only one fragment and that is the minimum spanning tree.
\end{enumerate}
Two main problems with these simple-looking steps are (1) {\it coordination}: the requirement that all nodes of a fragment coordinate their actions, i.e. they have to cooperate in order to find out the fragment's minimum outgoing edge, (2) {\it synchronization}: two nodes can be in the same fragment, but not be aware of this fact yet. This will lead to forming cycles.\\	
One of the major innovations of the paper ~\cite{Gallager} was the concept of a level. Levels characterize fragments and are defined as follows:
\begin{equation}
\begin{array}{l}
0\ if\ the\ fragment\ contains\ a\ single\ node \\ (L+1)\ if\ two\ fragments\ of\ level\ L\ join \\ L_1\ if\ two\ fragments\ of\ level\ L_1\ and\ L_2\ join\ and\ L_1\ >\ L_2
\end{array}
\end{equation}
The edges on which the last join of fragments takes place becomes the {\it core} of the new fragment and the two nodes adjacent to that edge coordinate the action of finding the next {\it MOE} and joining with another fragment.
\subsubsection{Detailed description of the algorithm}
During the algorithm, a node is in one of the following states:
\begin{itemize}
\item {\it Sleeping}: the initial state
\item {\it Find}: while participating in a fragment's search for $MOE$.
\item {\it Found}: otherwise.
\end{itemize}
Each node classifies its adjacent edges to be in one of the following states:
\begin{itemize}
\item {\it Branch}: if the edge belongs to the $MST$ of the current fragment.
\item {\it Rejected}: if the edge is not a branch, and connects to another node in the same fragment.
\item {\it Basic}: otherwise, i.e. unexplored.
\end{itemize}
The algorithm uses the following messages:
\begin{itemize}
\item {\it Initiate(w, L, s)}: Sent by core nodes (nodes adjacent to the core edge) to nodes in the fragment, right after the creation of the fragment asking them to participate in the search for $MOE$. $w$ is the weight of the core, $L$ is the level, $s$ is the state of the core.
\item {\it Test(w, L)}: Sent by a node in state Find over its minimum Basic edge to the node at the other end of the edge to find out if it is an outgoing edge. $w$ is the weight of the core, $L$ is the level.
\item {\it Reject()}: Sent by a node as a response to a Test message, if it arrives from a node in the same fragment.
\item {\it Accept()}: Sent by a node as a response to a Test message, if it arrives from a node not in the same fragment.
\item {\it Report(w)}: Sent by a node $v$ to its parent $u$ in the spanning tree of the fragment during the search for $MOE$. $w$ is the weight of the local $MOE$ found by $v$.
\item {\it Change-core()}: Sent by the core nodes to the node adjacent to the new $MOE$ of the fragment found.
\item {\it Connect(w, L)}: Sent by the node adjacent to the $MOE$ of the fragment to the node on the other end of this edge, requesting  a connection (a join).
\end{itemize}
In the initial state, each node is in {\it Sleeping} state. A node that spontaneously wakes up, or is awakened by receiving a message is a fragment of level $0$, and is in {\it Find} state.
After the initial state, the later executions of the algorithm follow an iteration of the following two basic procedures:
\begin{itemize}
\item Finding Procedure
\item Joining Procedure
\end{itemize}
{\bf Finding Procedure}\\
\indent {\bf \it Step I: broadcasting initiate messages}
\begin{nonumalgorithm}
\     $\star$ The core nodes broadcast an {\it Initiate(w, L, Find)} message on the 
\         outward Branches where $(w,L)$ is the identity of the fragment.
\     $\star$ A node v on receiving an Initiate message does the following:
\           $\diamond$ Changes to Find state.
\           $\diamond$ Updates local information about its fragment: the core and 
\               the level.
\           $\diamond$ Records the direction towards the core (i.e. the edge on which 
\              it received the message) so as to create a "parent-child" 
\              hierarchy.
\            $\diamond$ Forwards the Initiate message on the outward Branches, if any.
\           $\diamond$ Starts the MOE search procedure.
\end{nonumalgorithm}
\indent {\bf \it Step II: Finding the local MOE}
\begin{nonumalgorithm}
\     $\star$ A node u in fragment $F_1$ with id $(w_1,L_1)$ picks its 
\       minimum Basic edge, e, and sends on it a $Test(w_1,L_1)$ message.
\     $\star$ A node v in fragment $F_2$ with id ($w_2$ ,$L_2$) on receiving 
\       a $Test$ message does the following:
\         $\diamond$ If ($w_1$,$L_1$) == ($w_2$ ,$L_2$), e is not an outgoing 
\           edge. v sends a Reject() message to u; both u and v mark e as 
\           $Rejected$. $u$ goes to step 1.
\         $\diamond$ If $(w_1,L_1)\ \neq\ (w_2, L_2)$ and $L_2\ \geq \  L_1$, 
\           $v$ sends Accept() message to u. u marks e as its local MOE.
\         $\diamond$ If $(w_1,L_1)\ \neq\ (w_2 ,L_2)$ and $L_2\ <\ L_1$, 
\             $v$ does not reply to $u$'s message, until one of the above conditions 
\            is satisfied. This blocks $u$, since $u$ does not send a Report
\            message until it gets a reply for its Test message. This also
\            blocks the whole process of finding the $MOE$ of $F_1$.
\end{nonumalgorithm}
\indent {\bf \it Step III: Reporting the local MOE and deciding the MOE of the fragment}
\begin{nonumalgorithm}
\     $\star$ A leaf node sends Report(w) to its parent node, where $w$ is 
\        weight of local MOE. If the node has no outward Branches, then it
\        sends $Report(\infty)$
\     $\star$ An internal node u waits till it receives $Report$ messages on all
\        its outward Branches, finds the minimum weight, $w$, among them
\        including weight of its own local $MOE$, and sends a $Report(w)$ 
\        to its parent. 
\        If $w$ was received in a $Report$ message on edge $e$, it marks $e$ as 
\        the best edge.
\     $\star$ The core nodes decide which edge is the $MOE$. The core node
\         sends a $Change-core()$ message along the path of best edges, 
\         till it reaches the chosen node, which does not have a best edge.
\         Along the path, the "parent-child" pointers get reversed.
\     $\star$ The chosen node sends a $Connect(w,L)$ message over its $MOE$, 
\         and denotes the edge as a $Branch$.
\end{nonumalgorithm}
\indent {\bf Joining Procedure}\\
Suppose $u$ on fragment $F_1 (V_1,E_1)$ with id $(w_1,L_1)$ sends a $Connect$ message to node $v$ on fragment $F_2 (V_2 ,E_2)$ with id $(w_2 ,L_2)$ over edge $e$.
\begin{nonumalgorithm} 
\     $\star$ If $L_2$ == $L_1$ and $v$ is going to send, or has already sent, a 
\       $Connect$ message to $u$ on $e$, then combination takes place. A new 
\       fragment $F$ with nodes $V_1 \cup  V_2$  and edges $E_1 \cup E_2 \cup \{e\}$ 
\       is created. Level of $F$ is $L_1\ +\ 1$, core is $e$. Now the core 
\       nodes of the new fragment initiate another phase by sending an 
\       $Initiate(w(e), L_1 + 1, Find)$ message. In case $L_2$ == $L_1$, but 
\       the $MOE$s of $F_1$ and $F_2$ are different, $u$ waits till one of 
\       the conditions is satisfied.
\     $\star$ If $L_2\  >\  L_1$, then absorption of $F_1$ into $F_2$ takes place. 
\       Level of the expanded fragment is still $L_2$, and its core is the
\       core of $F_2$ .
\     $\star$ The situation in which $L_2\ <\ L_1$ is impossible, since a $Connect$ 
\     message is never sent in such a case.
\end{nonumalgorithm}
{\bf Correctness Proof}
\begin{lemma}
The algorithm is deadlock free i.e. they do not create any cycle.
\end{lemma}
{\bf Proof:} Decisions within a fragment are taken in a centralized way:it decides to join to one fragment at a time. 
\begin{figure}[ht]
\begin{center}
\vskip12pt
\epsfbox{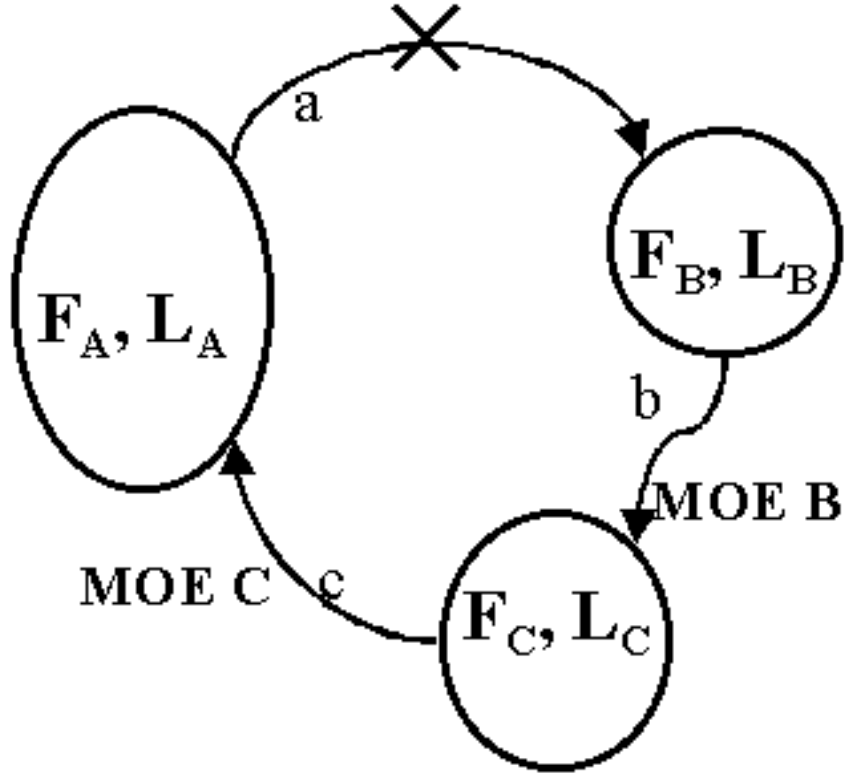}
\caption{GHS algorithm can't create cycles} \label{nocycle}
\end{center}
\end{figure}

Assume a cycle of fragments exists as shown in figure ~ref{nocycle}. It is obvious that a cycle would have to include an absorption, because combination is done only along common MOEs for both fragments. Assuming an absorption, say $F_C$ to $F_A$, assume $L_A > L_C$. Assume $F_B$ wants to join with $F_C$, then $L_B\ \geq \ L_C$. Obviously, $L_A\ >\ L_B$, and thus $F_A$ can neither do a combination or an absorption with $F_B$.\\
\par \indent {\bf Communication Complexity}
\begin{lemma} \label{levelfrag}
A fragment of level L contains at least $2^L$ nodes.
\end{lemma}
{\bf proof}: The proof is by induction on L.
Base case: Straightforward! A fragment of level 0 contains a single node. (by definition)
Induction step: Assume the lemma holds for fragments of levels $ \leq \ (L-1)$.
Consider a fragment $F$ of level $L$. $F$ was created by combining two fragments of level (L-1) and perhaps absorbing some smaller fragments. 
By the induction hypothesis, each one of the L-1 level fragments contains at least $2^{(L-1)}$ nodes. Thus F contains at least $(2^{(L-1)} +  2^{(L-1)})$ = $2^L$ nodes.
\begin{theorem} \label{upperlevel}
$\log N$ is an upper bound on fragment levels.
\end{theorem}
{\bf Proof:} Follows from the above lemma ~ref{levelfrag}.
\begin{theorem}
Message Complexity of the algorithm is $O(E + N \log N)$
\end{theorem}
{\bf Proof:} 
\begin{enumerate}
\item Each edge is rejected only once, and each rejection requires two messages (two test messages or a test and a reject message). So there are at most $2E$ messages leading to edge rejection.
\item At each level, the number of messages that a node receives or transmits is bounded by a constant:
\begin{enumerate}
\item At each level, a node can receive at most one Initiate message and one Accept message.
\item It can transmit at most one successful Test message, one Report message, and one Change-core or Connect message.
\item Thus, each level introduces 5N messages.
\end{enumerate}
\item According to the above theorem, maximum level is $\log N$ $\Rightarrow$ total number of other messages is $5N \log N$.
\item So communication complexity = $O(E + N \log N)$
\end{enumerate}
\par \indent {\bf Time Complexity}\\
Assume: All processors start simultaneously. Messages are delayed exactly one time unit. The algorithm is executed in rounds, as if the system is synchronous.
\begin{lemma} \label{timeunit}
It takes at most $(5lN - 3N)$ time units until all nodes are at level $l$.
\end{lemma}
{\bf Proof:} By induction on level number.\\
\begin{itemize}
\item Base case: l = 1. To wake up all nodes originally, it takes at most (N-1) time units. By time N, each node sends a Connect message. By time 2N, each node must be at level 1 through the propagation of Initiate messages. (Imagine processors arranged in a straight line with decreasing weights as we go down the line!!).
\item Induction Step: Assume the lemma holds good for level l.\\
At level l, each node can send at most N Test messages. At the worst case, these will be answered before time $(5lN - N)$. (Imagine each node except the last sends a Reject message (time N), last node delays responding till its level increases to l (time $5lN - 3N$ by induction step.) and it receives an Initiate message updating its local information about level (time N). Total time = $N + 5lN - 3N + N = 5lN - N).)$\\
Propagation of Report to Core, Change-core and Connect, and Initiate messages can take at most 3N time units.
Total time = $5lN - N + 3N$ = $5lN + 5N - 3N$ = $5(l+1)N - 3N$. Hence proved.
\end{itemize}

\begin{theorem}
Time Complexity is $O(N \log N)$.
\end{theorem}
{\bf Proof:} Follows from lemma ~\ref{timeunit} and theorem ~\ref{upperlevel}.\\

\par \indent {\bf Tightness of time complexity bound}\\
\begin{figure}[ht]
\begin{center}
\vskip12pt
\epsfbox{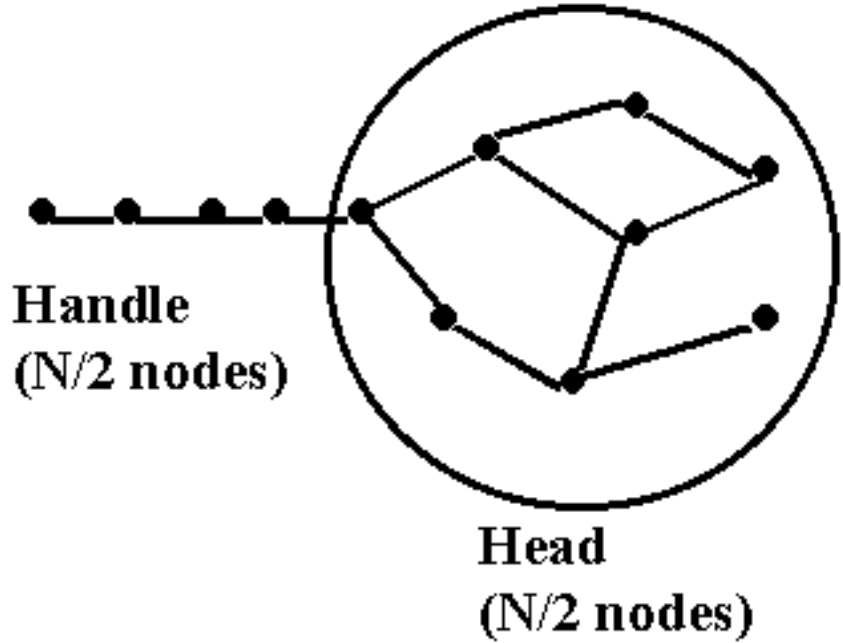}
\caption{Tightness of time complexity bound} \label{worstcase}
\end{center}
\end{figure}
We can prove the upper bound by the following example ~\ref{worstcase}. Edge weights in the handle increase as one gets away from the head: all nodes in the handle will be in the same fragment at level 1. If processing the head requires $(\log N)/2$ levels, and at each level, one fragment joins the handle, time = $\Theta (N \log N)$
\subsection{The node counting improvement ~\cite{Chin}}
The major innovation of the algorithm is that it tries to keep the fragment level a better estimate of the fragment size. It is obvious that nay fragment of level $L$ must have at least $2^L$ nodes. However, this is just a lower bound of its size: the fragment may have many more nodes than $2^L$ if it has accepted
a lot of submissions. The modified algorithm of Chin and Ting ~\cite{Chin} demands that
\begin{equation} \label{sizeequality}
2^L\ \leq\ size(F)\ <\ 2^{L+1}
\end{equation}
Tracking the fragment size can be achieved by having the root count the $report$ messages it receives. More accurately, each $report$ message has a counter that is increased at each hop of the message. Each node adds the counters of all the messages that it receives.
At the root, the level of the fragment is compared with the size. If $size(F)\ \geq\ 2^{L+1}$, then the level is increased till it satisfies ~\ref{sizeequality}. Then an $initiate$ message is broadcasted and
the procedure of finding the $MOE$ is repeated. This procedure is called {\it Root Level Increase}.
\par \indent This procedure increases the efficiency of the algorithm to $\Theta(n.g(n))$ where $g(n)$ is the number of time the $\log$ function must be applied to $n$ to get a value less than or equal to $1$. 
\subsection{The optimal algorithm of Awerbuch ~\cite{Awerbuch}}
This is the first algorithm that achieved optimal bounds for both communication and time. The algorithm is divided into phases and parts as follows:
\begin{enumerate}
\item {\it Counting Nodes Phase:} In this auxiliary phase, the nodes of the network are counted and a spanning tree is found that helps in counting. Weights are neglected and the joining policy is changed so that each fragment joins along the edge that leads to a greater fragment. The communication and time complexity
 of this phase are $O(m+n \log (n))$ and $O(n)$ respectively. Having a spanning tree, the number of nodes in the network can be counted.
 \item {\it MST Phase:} This phase is where the $MST$ is determined. It is divided into two parts:
\begin{itemize}
\item {\it Fragment's size:} $0$ to $\frac{n}{\log (n)}$. In this part, the algorithm behaves exactly the same as GHS algorithm ~\cite{Gallager}. The complexity remains optimal because the algorithm ends when the sizes of fragments become $\frac{n}{\log (n)}$.
 \item {\it Fragment's size:} $\frac{n}{\log (n)}$ to $n$. Here two new procedures are brought into action. (1) {\it Root Update} procedure. This resmbles the Root Level Increase procedure described in the previous section. The difference is that instead of counting the number of $report$ messages,
 the existence of "long" paths is detected. The $initiate$ message has a counter which is initialized to $2^{L+1}$ and is decreased at each hop. When the counter becomes negative, a message is sent back to the root. The level is increased by $1$ and a new $initiate$ message is issued. (2) {\it Test Distance} procedure. This procedure
  applies to fragments that have just submitted. The fragment tests to see if its distance from the new root is big enough to justify a level increase. Thus instead of staying idle, it manages to have its level increased in time related to the level.
 \end{itemize}
\end{enumerate}
\par \indent It is worth mentioning that there are some cases where Awerbuch's optimal algorithm ~\cite{Awerbuch} can create cycles or fails to achieve optimal time complexity. This fact has been shown in a paper by Faloutsos ~\cite{Faloutsos}. They have shown how to modify the
algorithm to avoid these problems and demonstrate both the correctness and optimality of the revised algorithm.
\section{Sequential Dynamic MST Algorithms}
There are two fully dynamic data structures for the general problem of dynamic graph algorithms as defined in section ~\ref{introduction}. They are (1) the dynamic trees of Sleator and Tarjan ~\cite{Sleator}, and (2) the topology trees of Frederickson ~\cite{Frederickson}. Both data structures follow the common principles of partioning the tree into a set of vertex-disjoint fragments, and making the least amount of modifications to maintain those partions in case of graph changes. However, they are different in how this partition is chosen. We would like to mention here that the dynamic trees of Sleator and Tarjan ~\cite{Sleator} are more suited towards simpler dynamic graph problems like the dynamic expression trees, compression and expansion of edges etc. In fact, they can be used to solve the problem of maintaining $MST$ of a graph when cost of a nontree edge $(v,w)$ decreases in $O(\log n)$ time. On the other hand, the topology trees of Frederickson ~\cite{Frederickson} are used as the basic building blocks of dynamic graph problems, and in fact, we will use the same idea while giving one of our distributed dynamic MST algorithm. So we will study the topology trees in more detail.
\indent The dynamic trees of Sleator and Tarjan ~\cite{Sleator} are able to maintain a collection of rooted trees, each of whose edges has a real-valued cost, under an arbitrary sequence of the following operations:
\begin{tabbing}
\ {\it maketree(v):}\  \= initialize a new tree consisting of a single vertex \\
\> $v$ with cost $0$.\\
\ {\it findroot(v):}\ return the root of the tree containing vertex $v$. \\
\ {\it findcost(v):}\ return a vertex of minimum cost in the path from $v$ to \\
\> $findroot(v)$.\\
\ {\it addcost(v, $\delta$):}\ \= add the real number $\delta$ to the cost of \\
\> every edge in the path from $v$ to $findroot(v)$.\\
\ {\it link(v,w):}\ \= Merge the trees containing vertices $v$ and $w$ by \\ 
\> inserting edge $(v,w)$.\\
\ {\it cut(v):}\ \= delete the edge leaving $v$, thus splitting into two the tree\\
\> containing vertex $v$.\\
\ {\it evert(v):}\ make $v$ the root of its tree.
\end{tabbing}
\begin{theorem}
Each of the above operations can be supported in $O(\log n)$ worst-case time.
\end{theorem}
{\bf Proof:} See ~\cite{Sleator}.\\
Given these operations, we can solve the case of a nontree edge decreasing weight through the series of operations: $findcost(v)$, $cut(v)$, $link(v,w)$.
\subsection{Clustering and Topology Trees}
Let $G(V,E)$ be a graph, with a designated spanning tree $M$. $Clustering$ is a method of partioning the vertex set $V$, into connected subtrees in $M$, so that each subtree is only adjacent to a few other subtrees. Before proceeding further, it is necessary to mention here that Frederickson's techniques use graphs in which no vertex has degree greater than $3$. The paper also provides a transformation from any graph to such a structure ~\cite{Frederickson}.
A {\it vertex cluster} with respect to the tree $M$ is a set of vertices that induces
a connected subgraph on $M$. An edge is $incident$ to a cluster if exactly one of its
end points is inside that cluster. Two clusters are $adjacent$ if there is a tree
edge that is incident to both. A {\it restricted partition of order z} of G is a partition of
its vertex set $V$ into $O(m/z)$ vertex clusters such that:
\begin{enumerate} \label{restrictedpartition}
\item Each set in the partition yields a vertex cluster of external degree at most $3$.
\item Each cluster of external degree $3$ is of cardinality $1$.
\item Each cluster of external degree less that $3$ is of cardinality less than or equal to $z$.
\item No two adjacent clusters can be combined and still satisfy the above.
\end{enumerate}
A restricted partition of order z can be found in linear time ~\cite{Frederickson}.
We now discuss how to update the clusters of a restricted partition of order $z$ when
the underlying graph is subject to updates. The basic update is a $swap(e,f)$: that is
replace a tree edge $e$ by a non-tree edge $f$, yielding a new spanning tree. This is
a basic update operation, because each edge insertion, deletion, and edge cost change causes
at most one swap in a spanning tree. We do the following to maintain the clusters:
\begin{itemize}
\item remove $e$. It splits $M$ into two trees $M_1$, and $M_2$. $M_1$ and $M_2$ inherit
all the clusters of $M$ and the following cases arise:
\begin{itemize}
\item if $e$ is entired contained in a cluster, that cluster is no long connected and therefore must be split.
After the split, we must check whether each of the two resulting clusters can be merged with neighboring clusters
in order to maintain codition (4) above.
\item if $e$ is between two clusters, then no split is needed. However, since the tree degrees of the clusters containing the endpoints of $e$ have been decreased, we must check if each cluster should be combined with an adjacent cluster, again because of condition (3).  
\end{itemize}
\item add $f$. $M$ inherits all clusters from $M_1$ and $M_2$, and the following cases arise:
\begin{itemize}
\item $f$ increases the tree degree of a cluster from $1$ to $2$. In order to preserve condition (3) above, we must check if this cluster must be combined with the cluster newly adjacent to it.
\item $f$ increases the tree degree of a cluster containing more than one vertex from $2$ to $3$. In order to satisfy condition (1), we have to split the cluster. After splitting, we have to again check if each cluster must be combined with an adjacent cluster. 
\end{itemize}
\end{itemize}
A {\it restricted multi-level partition} consists of a collection of restricted partitions of $V$ satisfying the following:
\begin{enumerate}
\item The clusters at level $0$ (known as {\it basic clusters}) contain one vertex each.
\item The clusters at level $l\ \geq\ 1$ form a restricted partition with respect to the tree obtained after shrinking all the clusters at level $l-1$.
\item There is exactly one vertex cluster at the topmost level.
\end{enumerate}
From the above definition, it follows that any cluster at level $l\ \geq\ 1$ is either (a) the union of two adjacent clusters of level $(l-1)$ such that the external degree of one cluster is $1$ or the external degree of both clusters is $2$, or (b) one cluster at level $(l-1)$. 
The {\it topology tree} is a hierarchical representation of $M$. Each level of the topology tree partitions the vertices of $M$ into connected subsets called {\it clusters}. More precisely, given a restricted multi-level partition for $M$, a {\it topology tree} for $M$ is a tree satisfying the following:
\begin{enumerate}
\item A topology tree node at level $l$ represents a vertex cluster at level $l$ in the restricted multi-level partition.
\item A node at level $l\ \geq\ 1$ has at most two children, representing the vertex clusters at level $l-1$ whose union gives the vertex cluster the node represents.
\end{enumerate}
\begin{theorem}
The update of a topology tree because of an edge swap can be supported in time $O(z+\log n)$. 
\end{theorem}
{\bf Proof:} For a proper proof of the theorem, see ~\cite{Frederickson}. We will give a brief idea of the proof. The update of a topology tree because of an edge swap in $T$ consists of two subtasks. First, a constant number of basic clusters (corresponding to leaves in the topology tree) have to be examined, and possibly updated. ($O(z)$). Second, the changes in these basic clusters percolate up in the topology tree, possibly causing vertex clusters in the multi-level partition to be regrouped in different ways. This involves a constant amount of work on at most $O(\log n)$ topology tree nodes.  \\
A {\it 2-dimensional topology tree} for a topology tree is defined as follows: For every pair of nodes $V_{\alpha}$ and $V_{\beta}$ at the same level in the topology tree, there is a node labeled $V_{\alpha} \times V_{\beta}$ in the {\it 2-dimensional topology tree}. Let $E_M$ be the tree edges of $G$. A node $V_{\alpha} \times V_{\beta}$ represents all the nontree edges of $G$(i.e. the edges of $E \setminus E_M$) having one end point in $V_{\alpha}$ and the other in $V_{\beta}$. The root of the {\it 2-dimensional topology tree} is labeled $V \times V$ and represents all the non-tree edges of $G$. If a node is labeled $V_{\alpha} \times V_{\beta}$, and $V_{\alpha}$ has children $V_{\alpha_i},\ 1\ \leq\ i\ \leq\ p$, and $V_{\beta}$ has children $V_{\beta_j},\ 1\ \leq\ j\ \leq\ q$, in the topology tree, then $V_{\alpha} \times V_{\beta}$ has children $V_{\alpha_i} \times V_{\beta_j},\ 1\ \leq\ i\ \leq\ p,\ 1\ \leq\ j\ \leq\ q$, in the {\it 2-dimensional topology tree}.
\par \indent Note that a {\it 2-dimensional topology tree} corresponds roughly to having $O(m/z)$ topology trees, one for each basic cluster in the restricted multi-level partition. As previously described, updating the basic clusters because of an edge swap would require a total of $O(z)$ time, and then updating these $O(m/z)$ topology trees would require a total of $O((m/z)\log n)$ time. This yields a total of $O(z+(m/z)\log n)$ time. The computational saving of a {\it 2-dimensional topology tree} is that it can be updated during a swap in its corresponding topology tree in $O(m/z)$ time only ~\cite{Frederickson}. This leads to the following theorem:
\begin{theorem}
The update of a 2-dimensional topology tree because of an edge swap in the corresponding topology tree can be supported in time $O(m/z)$.
\end{theorem}
{\bf Proof:} See ~\cite{Frederickson}. \\
Typical algorithms will balance this bound by choosing $z\ =\ \Theta(m^{1/2})$ to get an $O(m^{1/2})$ total time bound.
\begin{theorem}
The minimum spanning tree of an undirected graph can be maintained in time $O(m^{1/2})$ per update, where $m$ is the current number of edges in the graph.
\end{theorem}
{\bf Proof:} We maintain a restricted multi-level partition of order $z$, and the corresponding
topology tree and 2-dimensional topology tree as described before. We augment the 2-dimensional topology tree as follows:
Each leaf $V_i \times V_j$ stores the set $E_{i,j}$ of edges having one endpoint in $V_i$ and the other in $V_j$, as well as the minimum cost edge in this set.
This information is stored in a heap-like fashion: internal nodes of the 2-dimensional topology tree have the minimum of the values of their children.
This additional information required constant time per node to be maintained. Consequently, the update of this augmented 2-dimensional topology tree because of a swap can be done in $O(m/z)$ time.
\par \indent Whenever a new edge is inserted or nontree edge has its cost decreased, we can find a replacement edge is tie $O(\log n)$ with the dynamic trees of Sleator and Tarjan ~\cite{Sleator}. Whenever
an edge is deleted, or a tree edge has its cost increased, we can find a replacement edge as follows: let $e$ be the edge that has been deleted or increased. We first split the 2-dimensional topology tree at $e$ in $O(z+m/z)$ time.
Suppose this splits the corresponding topology tree into two trees, whose roots are the clusters $V_{\alpha}$ and $V_{\beta}$, with $V_{\beta}$ having no fewer levels than $V_{\alpha}$. To find a possible replacement edge for $e$, we examine the values at the nodes $V_{\alpha} \times V_{\gamma}$ for all possible $V_{\gamma}$ in the 2-dimensional topology tree, and take the minimum. It takes $O(m/z)$ time to find and examine those nodes.
\par \indent This yields a total of $O(z+(m/z))$ time for each update. Choosing $z\ =\ m^{1/2}$ gives an $O(m^{1/2})$ bound. However $m$ changes because of insertions and deletions. When the value of $z$ changes because of insertions and deletions. When the value of $z$ changes because of $m$, there will be at least
$m^{1/2}$ update before $z$ advances to the next value up or down in the same directions. Since there are at most $O(m/z)$ basic clusters that need to be adjusted, we can adjust a constant number of clusters during each update.
\subsection{Sparsification}
Sparsification is a generic technique for designing dynamic graph algorithms, due to Eppstein ~\cite{Eppstein}. It can be used to speed up many fully dynamic graph problems. Roughly speaking, when the technique is applied, it speeds up
a $T(n,m)$ time bound for a graph with $n$ vertices and $m$ edges to $T(m,O(n))$, i.e. to the time needed if the graph were sparse. E.g. if $T(n,m)\ =\ O(m^{1/2})$, we get a better bound of $O(n^{1/2})$ by applying this "black box" technique.
\par \indent The technique itself is quite simple. Let $G$ be a graph with $m$ edges and $n$ vertices. We partition the edges in $G$ into a collection of $O(m/n)$ sparse subgraphs, i.e. subgraphs with $n$ vertices and $O(n)$ edges. The information relevant for
each subgraph can be summarized in an even sparser subgraph, which is called as a {\it sparse certificate}. We merge certificates in pair, producing large subgraphs which are made sparse by again computing their certificate. The result is a balanced binary tree in which each node is represented by a sparse certificate. Each update involves $\log (m/n)$ graphs with $O(n)$ edges each, instead of one graph with $m$ edges. With some extra care, the $O(\log (m/n))$ overhead term can also be
eliminated ~\cite{Eppstein}. 
\subsection{Our Sequential Dynamic MST Algorithm}
We consider the problem of maintaining a minimum spanning tree during an arbitrary sequence of edge insertions and deletions. Given an $n$-vertex graph $G$ with edge weights, the fully dynamic minimum spanning tree problem is to maintain a minimum spanning tree $T$ under an arbitrary sequence of the following update operations: \\
\hspace*{0.35in} {\it increase}(e, $\delta$): Add the real number $\delta$ to the weight of the graph edge \emph{e = (u,v)} of $G$. \\
\hspace*{0.35in} {\it decrease}(e, $\delta$): Subtract the real number $\delta$ from the weight of the graph edge \emph{e = (u,v)} of $G$. \\
It's worth noticing that structural changes to $G$ like insertion of an additional edge or deletion of an existing edge can be modelled by the above two operations by doing the following: \\
\hspace*{0.35in} Whenever an edge is deleted, perform the operation: increase(e, $ \infty $). As any other edge in the graph would have less weight, this edge would disappear from the MST if it was there before. \\
\hspace*{0.35in} Whenever an edge is inserted, consider it as though the edge existed in the graph with a weight of $ \infty $ and now it has decreased its weight to $w$.
\subsubsection{Preliminaries}
There are several cases to be handled in edge-cost updating:
\begin{enumerate}
\item Cost of a tree edge increases.
\item Cost of a non-tree edge increases.
\item Cost of a tree edge decreases.
\item Cost of a non-tree edge decreases.
\end{enumerate}
Clearly, for the cases 2 \& 3, there will be no change in the minimum spanning tree. In the remaining two cases, the minimum spanning tree may change : one non-tree edge may replace one tree edge. These cases may be detected as follows:\\
\hspace*{0.35in} If the cost of a nontree edge \emph{e=(u,v)} is decreased, determine if the maximum cost of an edge in the cycle that \emph{e} induces in the tree has greater cost than cost of \emph{e}. If it has, then that edge will be replaced by {\it e}. An obvious implementation of this test would use $\Theta(n)$, becase there can at most be (n-1) edges in the cycle that is connected in the tree.\\
\hspace*{0.35in} If the cost of a tree edge {\it e=(x,y)} is increased, determine if the minimum cost nontree edge {\it (u,v)} that connects the two subtrees created by removing {\it e} has cost less than the cost of {\it e}. If it has, then that nontree edge {\it (u,v)} will enter the tree, and {\it e} will be forced out of the tree. An obvious implementation of this case would test $\Omega(m)$ edges for a replacement.\\
\thispagestyle{plain}
\subsubsection{Previous Work}
\begin{tabular}{|l||l||l|l||l|l|l|}
\hline
 & &\multicolumn{2}{l|}{Sequential}&\multicolumn{3}{l|}{Distributed} \\
\cline{3-7}
 & Algo &Best & Worst & Best & Worst & Msg\\
 & & Case & Case & Case & Case & Complexity\\
\hline
 &Kruskal & $\Theta(m\log m)$ & $\Theta(m\log m)$ & ? & ? & ?\\
 &Prim & $O(m)$ & $\Theta(m\log n)$ & ? & ? & ?\\
Static & GHS & ? & ? & ? & $O(n\log n)$ & $O(e+n\log n)$\\
 &CT & ? & ? & ? & $O(n g(n))$ & $O(e+n\log n)$\\
 &Awe & ? & ? & $O(n)$ & $O(n)$ & $O(e+n\log n)$\\
\hline\hline
 &Frdksn & $O(\sqrt{m})$ & $O(\sqrt{m})$ & ? & ? & ?\\
Dynamic &Epp & $O(\sqrt{n})$ & $O(\sqrt{n})$ & ? & ? & ?\\
 & HK & $O(\log^{3} n)$ & $O(\log^{3} n)$ & ? & ? & ?\\
 & HT & $O(\log^{2} n)$ & $O(\log^{2} n)$ & ? & ? & ?\\
\hline
\end{tabular}
Note: $g(n)$ is the iterative logarithmic function i.e. $g(n)$ is the number of times $\log$ function must be applied to $n$ to get a result less than or equal to $1$.
\begin{tabular}{|p{10.7cm}|}
\hline
GHS - Gallager, Humblet, and Spira. \\CT - Chin and Ting. \\Awe - Awerbuch. \\Frdksn - Frederickson. \\Epp - Eppstein (applied a technique called "Sparsification" to Frdksn's algorithm). \\HK - Henzinger and King (This algorithm is randomized and the time complexity is amortized time per update).\\HT - Henzinger and Thourp (Also randomized, in fact an improvement over HK. Amortized time per update)\\
\hline
\end{tabular}
\subsubsection{The Algorithm}
As has been discussed earlier, in any change, at most one tree edge gets replaced by a non-tree edge. We leverage this fact and define a correspondence between the tree edges and non-tree edges. We also specify how to modify the correspondence when the edge replacement occurs i.e. at any point of time, we have a total function {\it f} so that if any tree edge $e1$ increases weight, we find out $e2 = f(e1)$, where $e2$ is a non-tree edge and replace $e1$ with $e2$ if $cost_{e2}^{old}$ $<$ $cost_{e1}^{new}$; similarly if any non-tree edge $e2$ decreases weight, we find out $e1 = f(e2)$, where $e1$ is a tree edge and replace $e1$ with $e2$ if $cost_{e2}^{new}$ $<$ $cost_{e1}^{old}$.
\subsubsection{Initialization}
\begin{tabbing}
\  Let \= $G$ = $(V,E)$\\
\> $T$ = set of all tree edges \\
\> $t$ = number of tree edges \\
\> $NT$ = set of all non-tree edges.\\
\> $nt$ = number of non-tree edges.\\
\> so that $V = T \cup NT$ and $m = t + nt$.\\
\> $C_{e_j}$ = \= set of all tree edges with which the non-tree edge $e_j$ \\ \> \> forms a cycle when $e_j$ is added to the MST.\\
\end{tabbing}
Arrange all $e_j$s in $NT$ in increasing order. Execute the following procedure:
\begin{algorithm}[Initialize]
{\bf begin}
\     $S = NT$. /*$S$ is the set of non-tree edges in increasing order of their weights*/
\     $W = \emptyset$. \nonumline \\{\it /*$W$ denotes the set of tree edges for which the responsibility set (which non-tree edge is responsible for each tree edge) has been found.*/}
\     {\bf while} there are edges in set S
\          $e_j$ = next edge from set S.  /*Next edge in sorted order*/
\          $S$ = $S$ $\setminus$ $\{e_j\}$.
\          calculate $U_{e_j}$ = $C_{e_j}$ $\setminus$ $W$. \nonumline \\{\it /*$U_{e_j}$ is the set of edges for which $e_j$ is responsible. These are the tree edges that are there in the cycle that it makes, but are not there among the set of tree edges for which responsibility has been found out. This is because we are going in a sorted order*/}
\          {\bf for} each edge $e_i$ in $U$ do 
\               set $f(e_i) = e_j$
\          {\bf end for} 
\          set $f(e_j)$ = $e_i$ where $e_i$ has the maximum cost 
\                          among the edges present in U.
\          $W = W \cup U$ {\it /*Update the set of tree edges for which responsibility set has been found out.}
\     {\bf end while} 
{\bf end}
\end{algorithm}

\subsubsection{Updation during structural change}
Whenever a change in edge weight occurs s.t. a tree edge is to be replaced, the new $C_{e_j}$s can be calculated as follows:

\begin{algorithm}[Updation]
Let $e_c$ be the non-tree edge to replace a tree edge.
\    Calculate $U = C_{e_c} \cup \{e_c\}$.
\    {\bf for} each $e_j$ in $NT$ do
\        $C_{e_j} = C_{e_j} \Delta U$.
\    {\bf end for}
\end{algorithm}
Here $\Delta$ stands for symmetric difference. After the cycle entries are calculated, the function can be recalculated in the same lines.
\subsubsection{Data Structures}
We will use a balanced tree (preferrably AVL tree) for our computation. Each node will represent a non-tree edge. Each nod ewill also contain the following additional information:
$C_{e_j}$ = the tree edges with which it makes a cycle.\\
$N_{e_j}$ = the tree edges for which it is responsible.\\
$L_{e_j}$ = the tree edges for which its left subtree is responsible.\\
$R_{e_j}$ = the tree edges for which its right subtree is responsible.\\

Initial building of the tree is very obvious. Now we will consider each of the two cases one by one:
\subsubsection{Non-tree edge decrease}
\begin{algorithm}
{\bf begin}
\       Search in the AVL tree for the node. [$\log(n)$ operation]
\       Get the maximum cost in N = $e_i$.
\       if $cost_{e_j}^{new} > cost_{e_i}^{old}$  then exit.
\       else 
\           /*the tree edge has to be replaced.*/
\           $C_{e_i}$ = $C_{e_j}$ - $e_i$ + $e_j$.
\           $N_{e_i}^{initial} = N_{e_j}$.
\           $L_{e_i}$ = $\emptyset$.    
\           $R_{e_i}$ = $\emptyset$.    
\           Traverse in the tree to insert the node starting from the root.
\           At the root $e_r$, initialize the following:
\              $L_{e_i}$ = $L_{e_r}$ if it has to go left, else 
\               initialize the $R_{e_i}$ = $R_{e_r}$.
\           At each of the nodes,
\           if it has to go left of node $e_k$, $L_{e_i}$ = $L_{e_i}$ - $L_{e_k}$.
\             else modify the R accordingly.
\           Each time, modify the N value also accordingly.
\           Do the insertion.

{\bf end}
\end{algorithm}
\subsubsection{tree edge increase}
This case also is the same as before except that during the search, we start from the root, search its L set, if it is there, we go left, else we go right. If the edge has to be replaced, the procedure is the same as the one given above.
\subsubsection{Time Complexity}
Insert, delete, and search take O($\log(n)$) time. Each time we are in a node, we are doing a constant number of set operations. If we can prove that these set operations also take O($\log(n)$) time, then the overall time complexity:\\
$T(n) = O(\log n)$
For the timebeing, I have found the set operations to be $O(n)$. So the time complexity is $O(n\log n)$.
\subsubsection{Example}
\begin{figure}[ht]
\begin{center}
\vskip12pt
\epsfxsize=5in
\epsfysize=4in
\epsfbox{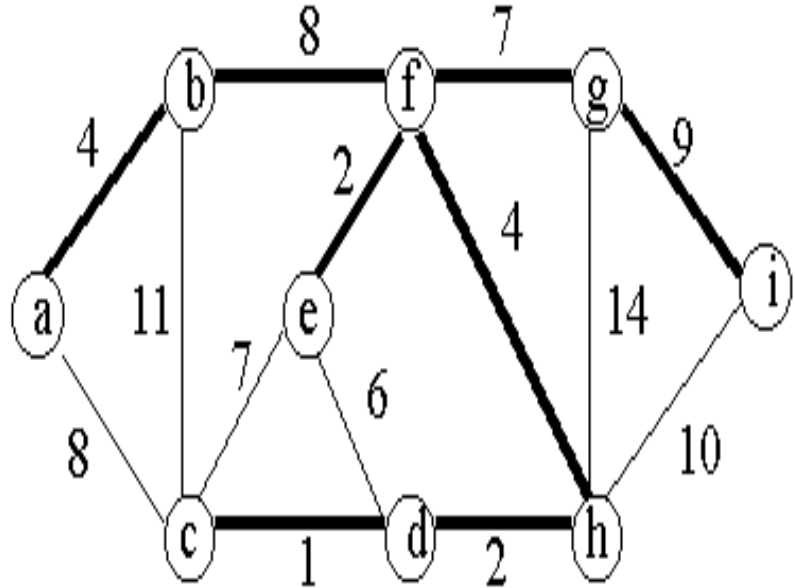}
\caption{A sample graph and its minimum spanning tree} \label{graph}
\end{center}
\end{figure}
\begin{figure}[ht]
\begin{center}
\vskip12pt
\epsfxsize=5in
\epsfysize=4in
\epsfbox{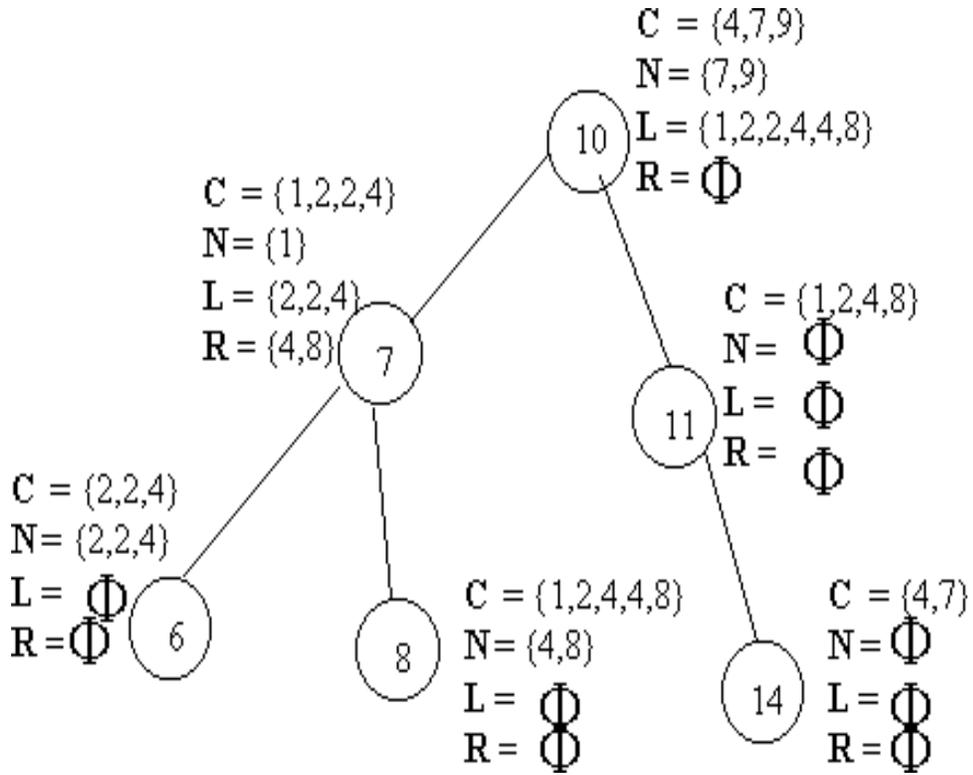}
\caption{Non-tree edges' AVL tree structure for the above graph} \label{avltree}
\end{center}
\end{figure}

The corresponding AVL tree of the non-tree edges and the sets for the nodes are given in figure~\ref{avltree}.
Now suppose edge with weight $4$ (in non-tree edge $6$'s cycle) increases weight to $9$. Search for $4$ in the tree. Start with root $10$'s $L$ vector. $4$ is there. Go left, you will eventually reach node $6$. New weight $8 > 6$. So that tree edge has to be replaced. Delete the node. Now you will have to insert $8$. Initially assume that $8$ is responsible for $\{2,6,2\}$ (i.e. the edges that $6$ was responsible for). Start at the root $10$. You go left. So you are still low enough to be responsible for all your edges. At node $7$, you have to go right. So surrender whatever nodes that node $7$ and its left subtree can be responsible for, because they have lower weight than you. This is clearly done by taking a symmetric difference at node $7$. Node $7$'s vectors become the following:
$C$ = $\{1,2,2,4\} \Delta \{2,2,4,6\} = \{1,6\}$.\\
$N$ = $\{1,6\}$ \\
Similarly, at node $8$, you surrender $2$. When node $9$ is inserted, it is left responsible for the other tree edge $2$.
\section{Distributed Dynamic MST Algorithms}
The only known parallel algorithm for updation of minimum spanning trees is due to Pawagi and Ramakrishnan~\cite{Pawagi}. They base the model of computation to be a parallel random access machine with multiple reads, but single writes. The algorithm described in the paper requires $O(\log n)$ time and $O(n^2)$ processors.
\subsection{Pawagi-Ramakrishnan's algorithm}
{\bf Definitions}\\
\indent Given a graph $G(V,E)$ and minimum spanning tree $M$, the {\it lowest common ancestor} of vertices $x$ and $y$ in $M$ is the vertex $z$ such that $z$ is a common ancestor of $x$ and $y$, and any other common ancestor of $x$ and $y$ in $M$ is also an ancestor of $z$ in $M$. \\
\indent An {\it inverted tree} is a tree where the edges of the tree $T$ are oriented from a vertex to its father. Note that the edges will be directed, and for each edge $(a,b)$ in the inverted tree, $b$ is the father of $a$. \\
\indent Let $T\ =\ (V',E')$ be an inverted tree with $V'\ =\ \{1,2,..,n\}$ and $r$ be the root of $T$. $T$ with a self-loop at its root $r$ represents a function $F:V' \mapsto V'$ such that $F(i)$ is the father of vertex $i$ in $T$ for $i \neq r$ and $F(r)\ =\ r$. From the function $F$, define $F^k,\ k\ \geq\ 0$, as follows:\\
\indent \indent $F^k:V' \mapsto V'\ (k\ \geq\ 0)$ such that
\begin{equation}
\begin{array}{l}
\forall\ i\ \in\ V',\ F^k(i)\ =\ i\ if\ k\ =\ 0.\\ \           F^k(i)\ =\ F(F^{k-1}(i))\ if\ k\ >\ 0.
\end{array}
\end{equation}
\begin{lemma}
Given the function $F$ of an inverted tree, the mapping $F^k$, $0\ \leq\ k\ <\ n$, can be computed in $O(\log n)$ time using $O(n^2)$ processors.
\end{lemma}
{\bf Proof:} see ~\cite{Pawagi}.
\begin{lemma}
We can compute the lowest common ancestors of $n \choose 2$ vertex pairs (number of unordered pairs of $n$ elements) in an inverted tree in $O(\log n)$ time using $O(n^2)$ processors.
\end{lemma}
{\bf Proof:} see ~\cite{Pawagi}.\\
\indent Let the function $MAX(e_1, e_2)$ return the maximum cost edge among the edges $e_1$ and $e_2$. Let $E_m^k(i),\ 1\ \leq\ i\ \leq\ n$ be the minimum cost edge on the path from $i$ to it $k^{th}$ ancestor in $T$. Then \\
\begin{equation}
\begin{array}{l}
E_m^1(i)\ is\ the\ edge\ (i, F^1(i)).\\E_m^k(i)\ is\ the\ edge\ MAX(E_m^{k-1}(i),(F^{k-1}(i),F^k(i))),\ k\ >\ 1.
\end{array}
\end{equation}
\begin{lemma}
The mapping $E_m^k(i)$, $1\ \leq\ i\ \leq\ n$, $0\ \leq\ k\ <\ n$, can be computed in $O(\log n)$ time using $O(n^2)$ processors.
\end{lemma}
{\bf Proof:} see ~\cite{Pawagi}.\\
Given these definitions and lemmas, the steps following to algorithms for maintaining minimum spanning trees become easy.
\begin{itemize}
\item Cost of a tree edge $(x,y)$ increases. We proceed as follows:
\begin{enumerate}
\item Delete the tree edge $(x,y)$. This step is to set $F^1(x)\ =\ x$.
\item Identify the vertices in each of these subtrees. This involves computing the function $F^k$.
\item Find the minimum cost edge $(u,v)$ connecting them. Essentially this is to compute the function $E_k^m$.
\item Add the edge $(u,v)$ to the forest.
\item Maintain the new $MST$ as an inverted tree. 
\end{enumerate}
\item Cost of a non-tree edge $(u,v)$ decreases. We proceed as follows:
\begin{enumerate}
\item Add $(u,v)$ to the old MST. It induces a cycle in the old MST. Calculate the new $F^k$ and $E_m^k$.
\item Remove the maximum cost edge from this cycle. Find out from the $E_m^k$ function computation.
\item Maintain the new $MST$ as an inverted tree.
\end{enumerate}
\end{itemize}
\begin{theorem}
Updation of a minimum spanning tree in this parallel computation requires $O(\log n)$ time and use $O(n^2)$ processors.
\end{theorem}
{\bf Proof:} Follows from the fact that each of the above steps takes $O(\log n)$ time using $O(n^2)$ processors (prove directly from the lemmas described).
\subsection{Our Distributed Dynamic MST Algorithm}
Our algorithm is based on the application of Frederickson's topology trees. We maintain partitions in a distributed computation and create topology trees from the clusters.
\begin{enumerate}
\item {\it Finding Clusters:} Given the initial $MST$, we can find the clusters by using the same idea as in the GHS algorithm ~\cite{Gallager} of forming fragments. However, here we will be bothered about
exploring only the tree edges, we ignore the weights and each fragment joins along an edge that leads to the greater fragment (to solve the synchronization problem). Each $report$ message contain a counter
in addition to other parameters, that is initialized to $0$, and each parent will sum up counters received from all its children before sending the $report$ message up.
Each $test$ and $accept$ message will contain the size of the current fragment and the decision of sending an $accept$ message will depend on whether
\begin{equation}
size(F\ sending\ test\ message)\ +\ size(current\ fragment)\ \leq\ z
\end{equation}
Otherwise, a $reject$ message is sent.
\item {\it Making the restricted partition of order z:} We assume that the graph $G$ has been transformed to form a graph of maximum vertex degree $3$. In order to satisfy the conditions in ~\ref{restrictedpartition}, we let every leaf of the
fragment send the number of tree edges incident to it through a $report$ message. Each parent sums up the number of incident tree edges for all its children and sends the $report$ message up. At the root (the core nodes), the number of incident tree edges is compared with the cardinality of the fragment to either split the fragment or let it join with other fragments.
\item {\it Forming the restricted multi-level partition:} This goes on the same way as the Frederickson's algorithm ~\cite{Frederickson}.
\end{enumerate}
After the 2-dimensional topology tree is formed, in response to an edge increase or decrease, we can efficiently find out the replacement edge by having each leaf $V_i \times V_j$ send the minimum of its $E_{ij}$ set to the parent. At the root, the minimum over all is taken.
Depending on the cost of the replacement edge, we can decide on whether to form a $swap(e,f)$. After the $swap$, further splitting might be necessary which can be done in ways similar to ones described above.
\section{Conclusion}
In this termpaper, we studied minimum spanning tree algorithms in both sequential and distributed case, with an aim to studying the dynamic algorithms for maintaining minimum spanning trees. Specifically, we studied Frederickson's topology trees ~\cite{Frederickson} in detail, which gives a time complexity of $O(m^{1/2})$, one the first best techniques for dynamic MSTs. We then discussed the
sparsification technique due to Eppstein ~\cite{Eppstein} which is a generic technique to speed up dynamic graph algorithms. We then presented a new approach of solving dynamic MSTs in sequential case by exploiting some new properties of the spanning trees. This algorithm has a best case complexity of $O(\log ^2)$. We also studied the dynamic MST problem in distributed system. To this regard, we gave a
brief idea of the parallel algorithm by Pawagi and Ramakrishnan ~\cite{Pawagi}. Based on the topology trees of Frederickson, we gave outline of how to create a distributed algorithm for solving the dynamic MST problem.
\par \indent In future, we indent to build upon the distributed algorithm given here and make a full-fledged algorithm. We also intend to study the sparsification technique in more detail and make it distributed, so that we can speed up the distributed algorithms further.

\end{article}

\begin{references}
\bibitem{Pawagi}
S. Pawagi and I.V. Ramakrishnan. An $O(\log n)$ algorithm for parallel update of minimum spanning trees. {\it Information Processing Letters}. Vol. 22. pp. 223--229. 1986.
\bibitem{Gallager}
R. G. Gallager, P. A. Humblet, and Philip M. Spira. A distributed algorithm for minimum-weight spanning trees. {\it ACM Transactions on
          Programming Languages and Systems}, Vol. 5, No. 1, pp. 66--77. January 1983.
\bibitem{Awerbuch}
Baruch Awerbuch. Optimal distributed algorithms for minimum weight spanning tree, counting, leader election and related problems. {\it In Proceedings of the Nineteenth Annual ACM Symposium on Theory of Computing}, pp. 230--240, New York City, 25-27
          May 1987.
\bibitem{Frederickson}
Greg N. Frederickson. Data structures for on-line updating of minimum spanning trees, with applications. {\it SIAM Journal on Computing}, Vol. 14, No. 4, pp. 781--798.
     November 1985.
\bibitem{Eppstein}
David Eppstein and Zvi Galil and Amnon Nissenzweig. Sparsification -- a technique for speeding up dynamic graph algorithms. {\it Journal of the ACM},
     Vol. 44, No. 5, pp. 669--696. September 1997.
\bibitem{CLR}
Thomas H. Cormen, Charles E. Leiserson, and Ronald L. Rivest. {\it Introduction to Algorithms}. MIT Press, Cambridge, MA. 1990.
\bibitem{Sleator}
D.D. Sleator and R.E. Tarjan. A data structure for dynamic trees. {\it Journal of Computer System Sciences}. Vol. 24, pp. 362--381. 1983. 
\bibitem{Attiya}
H. Attiya and J. Welch. {\it Distributed Computing:Fundamentals, Simulations and Advanced Topics}. McGraw Hill, London. 1998.
\bibitem{Chin}
F. Chin and H.F. Ting. An almost linear time and $(V \log V+E)$ messages distributed algorithm for minimum weight spanning trees. {\it Proceedings of the 1985 FOCS Conference}, Portland, Oregon. Oct 1985.
\bibitem{Faloutsos}
Michalis Faloutsos and Mart Molle. Optimal distributed algorithm for minimum spanning trees revisited. {\it Proceedings of Principles Of Distributed Computing (PODC)}. 1995.
\end{references}
\end{document}